\documentclass[aps,prl,twocolumn,superscriptaddress,amsmath,amssymb]{revtex4-1}
\usepackage{verbatim}
\usepackage{graphicx}
\usepackage{varioref}
\usepackage{hyperref}
\usepackage[capitalize]{cleveref}

\begin{document}

\title{Reservoir computing with a single time-delay autonomous Boolean node}

\date{\today}

\author{Nicholas D. Haynes}
\affiliation{Department of Physics, Duke University, Durham, North Carolina, USA}

\author{Miguel C. Soriano}
\affiliation{Instituto de F\'isica Interdisciplinar y Sistemas Complejos, IFISC (UIB-CSIC), Campus Universitat de les Illes Balears, Palma de Mallorca E-07122, Spain}

\author{David P. Rosin}
\affiliation{Department of Physics, Duke University, Durham, North Carolina, USA}

\author{Ingo Fischer}
\affiliation{Instituto de F\'isica Interdisciplinar y Sistemas Complejos, IFISC (UIB-CSIC), Campus Universitat de les Illes Balears, Palma de Mallorca E-07122, Spain}

\author{Daniel J. Gauthier}
\affiliation{Department of Physics, Duke University, Durham, North Carolina, USA}

\begin{abstract}

We demonstrate reservoir computing with a physical system using a single autonomous Boolean logic element with time-delay feedback. The system generates a chaotic transient with a window of consistency lasting between 30 and 300 ns, which we show is sufficient for reservoir computing. We then characterize the dependence of computational performance on system parameters to find the best operating point of the reservoir. When the best parameters are chosen, the reservoir is able to classify short input patterns with performance that decreases over time. In particular, we show that four distinct input patterns can be classified for 70 ns, even though the inputs are only provided to the reservoir for 7.5 ns.

\end{abstract}

\maketitle


Classical von Neumann machines cannot efficiently perform some tasks, such as pattern recognition and classification, that animal brains do with relative ease. This observation motivates the search for novel computational paradigms, particularly those with biological plausibility. One approach is reservoir computing (RC) \cite{Jae01,Maa02}, a method of efficiently training recurrent neural networks to perform a given task. In conventional RC, computation is performed by a large recurrent network of nonlinear elements with arbitrary connection weights (the reservoir) \cite{Luk09,Ham09}. The reservoir performs a nonlinear transformation on a time-dependent input signal, mapping it onto a high-dimensional state space in which it is linearly separable from other inputs. During training, the weights of connections within the reservoir are kept fixed, and the weights of connections to an output layer are determined by linear regression with target outputs.

Remarkably, a large, complex network is not necessary for RC. In fact, any system that is able to map input states onto a high-dimensional state space and that fulfills three commonly cited sufficient conditions for RC \cite{Maa02} -- separation of input states, generalization of similar inputs to similar outputs, and fading memory -- can be used \cite{App11}. For example, previous experiments \cite{App11,Lar12,Paq12,Sor13,Bru13,Dej14} have shown that a high-dimensional transient output can be generated with a single, physical nonlinear device that time-multiplexes ``virtual nodes'' using time-delay feedback. These experiments demonstrated that a RC with time-delay feedback can achieve excellent performance on time-dependent pattern recognition tasks such as speech recognition and time series prediction.

The success of these systems motivates the exploration of other physical time-delay systems that possess suitable characteristics for RC. In this vein, we introduce a RC consisting of a single autonomous Boolean logic element that executes the exclusive-or (XOR) function with two time-delay feedback lines. The state of an autonomous Boolean network with time-delay feedback depends on a continuous history of its past states, making the dynamics much more complex than in synchronous systems, which depend on only a discrete set of past values. This added complexity allows small, resource-efficient systems to perform the required high-dimensional transformation on inputs \cite{Ghi85,Ghi08}. 

Here, we demonstrate experimentally that this simple time-delay autonomous Boolean reservoir is capable of producing dynamics that fulfill the sufficient conditions for RC and can solve a classification task. We find that input signals produce chaotic transients with windows of consistency \cite{Uch04}. We then use a measure of effective dimensionality to show that optimal operating points of the reservoir can be found by varying the lengths of the time delays. Finally, we show that with an appropriate choice of parameters, the system is able to classify input signals with an accuracy that decreases with time.

\begin{figure}[htb]
	\centering
		\includegraphics[scale=1]{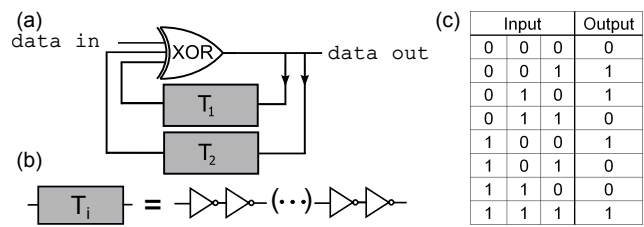}
	\caption{(a) Circuit diagram of the single-node time-delay reservoir. (b) Time delays are realized as a cascade of pairs of inverter gates. (c) Truth table for the three-input XOR gate. Two stable fixed points exist -- when all inputs are 0 (1), the output is 0 (1).}
	\label{fig1}
\end{figure}

%
%
We build the reservoir shown in \cref{fig1} on an Altera Cyclone IV field-programmable gate array (FPGA), an integrated circuit comprised of $\sim 10^5$ distinct logic elements. Each logic element can be configured to compute an arbitrary Boolean function on several inputs and wired to other logic elements on the chip using a hardware description language. In addition, the FPGA architecture allows for asynchronous logic operations, and the inherent rise time of the logic elements can be used to realize physical time delays.

The XOR operation is chosen as the Boolean node for several reasons. First, it exhibits maximum Boolean sensitivity -- when any of the inputs to a XOR gate are flipped, the output is also flipped (see \cref{fig1}c), which produces dynamics with maximal spreading with respect to the input states. Second, the XOR operation has a balanced output, meaning that, for the $2^n$ possible combinations of $n$-bit input patterns, half result in 0 and half result in 1. Choosing a balanced Boolean function prevents the system from being biased toward 0 or 1 over time. 

We realize the physical time-delay feedback lines by exploiting the finite response time of the FPGA's logic elements. Delay line 1 (2) is constructed by wiring $N_1$ ($N_2$) pairs of inverters in series; we call a pair of inverters a \textit{delay element}. A pair of inverters is used as a delay element instead of a single copier to correct for an observed asymmetry between the rise and fall times of the logic elements \cite{Ros13b}. When the lines are implemented on the FPGA, we observe a certain degree of heterogeneity in the delays due to physical imperfections. To reduce the effect of the heterogeneity as much as possible, we manually fix each of the delay lines to specific logic elements next to each other on the FPGA, rather than allowing them to be automatically placed by a compiler.

The experimentally measured delay times $T_1$ ($T_2$) resulting from $N_1$ ($N_2$) between 7 and 20 delay elements are listed in \cref{tab1}. The average delay time due to a single delay element is 0.59 ns, but the variation between elements in the same line is large, ranging between 0.43 ns and 0.99 ns. By manually fixing the lines, however, much of the heterogeneity between the two lines is eliminated -- the difference between $T_1$ and $T_2$ is, on average, 0.03 ns, and is never more than 0.1 ns.

\begin{table}[h]
\begin{center}
\caption{Measured delay times for each delay line.}
\begin{tabular}{|c|c|c|}
\hline
$N_1$, $N_2$ & $T_1$ (ns) & $T_2$ (ns)\\
\hline
7 & 3.76 & 3.75\\
\hline
8 & 4.31 & 4.31\\
\hline
9 & 5.30 & 5.25\\
\hline
10 & 5.79 & 5.79\\
\hline
11 & 6.35 & 6.44\\
\hline
12 & 6.79 & 6.89\\
\hline
13 & 7.50 & 7.49\\
\hline
\end{tabular}
\quad
\begin{tabular}{|c|c|c|}
\hline
$N_1$, $N_2$ & $T_1$ (ns) & $T_2$ (ns)\\
\hline
14 & 7.94 & 7.92\\
\hline
15 & 8.44 & 8.41\\
\hline
16 & 8.99 & 8.96\\
\hline
17 & 9.75 & 9.76\\
\hline
18 & 10.30 & 10.25\\
\hline
19 & 10.74 & 10.76\\
\hline
20 & 11.42 & 11.38\\
\hline
\end{tabular}
\end{center}
\label{tab1}
\end{table}

To demonstrate that this system fulfills the sufficient conditions for RC described above, we observe the transient response of the reservoir to different inputs. Input data are encoded in words of bits that are supplied to the reservoir in the form of sequential strings of little-endian (least-significant bit first), non-return-to-zero digital voltages at 400 MHz. After a word is completed, the \texttt{data in} wire of the reservoir is tied to 0. Similar initial conditions are assured by allowing the system to settle into the fixed point $\text{XOR}(0,0,0)=0$ before inputting data. A header to the input is necessary to distinguish patterns with leading or trailing zeros -- to a reservoir sitting in the fixed point $\text{XOR}(0,0,0)=0$, the word 0100, for example, appears to be just a time-shifted version of 0010. Including a one-bit header changes 0100 to 01001 and 0010 to 00101, which appear as distinct inputs.

\begin{figure}[htb]
	\centering
		\includegraphics[scale=1]{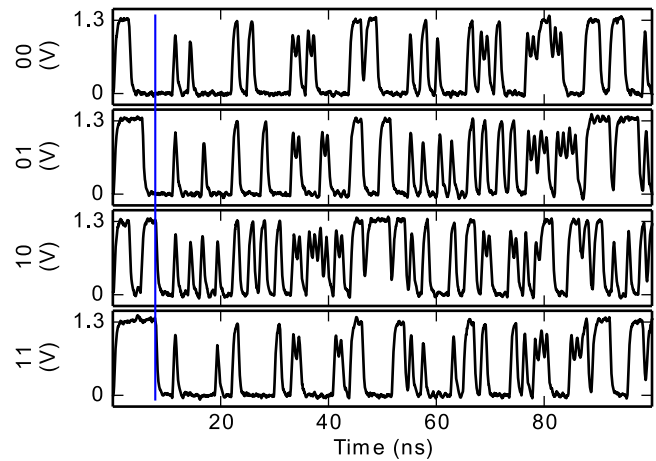}
	\caption{(color online) Experimentally observed transient dynamics of four unique input patterns (noted on the y axes), generated by a reservoir with $N_1 = 17$, $N_2 = 18$. Solid blue (dark gray) vertical line marks $t=7.5$ ns, where the input word ends.}
	\label{fig2}
\end{figure}

The circuit in \cref{fig1}a is predicted \cite{Cav10,Ghi85,Ghi08} to have aperiodic behavior when the time delays are incommensurate. Indeed, when data is fed to the experimental reservoir, a long, complex transient is produced. The first 100 ns of four complex transients captured by a high-speed oscilloscope is shown in \cref{fig2} for $N_1 = 17$, $N_2 = 18$ elements. Each of the transients is generated by a unique 2-bit input word. Clear Boolean-like transitions between high and low voltages can be observed, albeit with a finite slew rate. By direct inspection, it is clear that the output patterns resulting from each of the distinct inputs are unique. We find that a simple event-based model including a low-pass filter to remove short pulses can reproduce the observed dynamics \cite{Zha09,Cav10}, but it is not described further here.

In addition to output states that are unique to the inputs, reliable computation requires reproducibility of the outputs, \textit{i.e.}, reservoir responses to the input signals must be consistent \cite{Uch04}. Starting with the reservoir in its fixed point, we repeatedly observe transients by providing a word to the reservoir, recording the output time series for 1 $\mu$s, and then forcing the system back into its fixed point before repeating the experiment.

\begin{figure}[htb]
\centering
\includegraphics[scale=1]{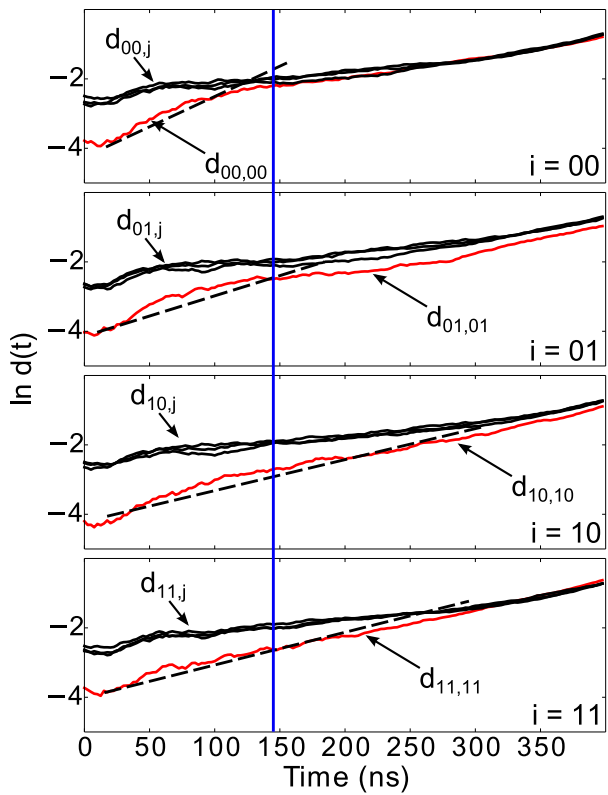}
\caption{(color online) Boolean distances ($\tau=100$ ns) $d_{i,i}$ and $d_{i,j}$ ($i \neq j$) averaged over 50 output time series from a reservoir with $N_1 = 8$, $N_2 = 11$ elements for four distinct inputs. Solid blue (dark gray) vertical line represents the end of the consistency window, beyond which states are not distinguishable.}
\label{fig3}
\end{figure}

To quantify the consistency of reservoir states, we define the measure of the output state space as the Boolean distance
\begin{equation}
d_{i,j}(t) = \frac{1}{\tau}\int_t^{t+\tau} x_i(t') \oplus x_j(t') dt' ,
\end{equation}
where $x_i(t) \in \{0,1\}$ is the Booleanized time-dependent output state corresponding to input $i$. We inject the four 2-bit input patterns ($i \in \{00, 01, 10, 11\}$) into a reservoir with $N_1 = 8$, $N_2 = 11$ elements. The average distances between the output time series from each of the input words is shown in \cref{fig3}. Consistency can be observed by comparing the distances $d_{i,i}(t)$ between transients generated by the same inputs with those generated by different inputs ($d_{i,j}(t)$, $i \neq j$). For each $i$, $d_{i,i}(t=0)$ is significantly smaller than $d_{i,j}(t=0)$, but increases exponentially until it converges with $d_{i,j}(t)$. The window of exponential divergence in $d_{i,i}(t)$, a signature of a chaotic transient, is indicated by dashed black lines in \cref{fig3}. The slopes of these lines, which indicate how quickly $d_{i,i}(t)$ diverges, correspond to the local Lyapunov exponents of the system.

The minimum time to the convergence of $d_{i,i}(t)$ with $d_{i,j}(t)$ in \cref{fig3} is approximately 145 ns, and thus the transient $x_i(t)$ due to input $i$ is distinguishable from $x_j(t)$ for 145 ns. We call this period of time the consistency window. For $T_1$, $T_2$ between 3.75 and 11.42 ns ($N_1$, $N_2$ between 7 and 20 elements), we find consistency windows between 30 ns and 300 ns. The existence of the consistency window suggests that the reservoir fulfills the separation and generalization requirements introduced above.

%
%
To quantify the ability of the reservoir to perform effective computation and characterize its performance as a function of the size of the delays, we consider the measures known as kernel quality and generalization ability \cite{Leg07,Bus10}. Once the consistency window for particular values of $N_1$ and $N_2$ is measured, the $m \times m$ state matrix $M_K$ is created using the following procedure. Data is collected from the reservoir by sampling the \texttt{data out} wire at 400 MHz to RAM blocks located on the FPGA. The size $m$ of the matrix is defined as the number of samples that can be collected within the consistency window ($12 \leq m \leq 120$ samples for consistency windows of $30 \leq L \leq 300$ ns). The state matrix $M_{K}$ is then formed by collecting $m$ samples from the resulting time series of $m$ distinct inputs. The normalized kernel quality $K = \frac{1}{m} \text{rank} \ M_{K}$ is defined as the rank of the normalized state matrix. A system with perfect separation will produce linearly independent outputs for all possible input states, \textit{i.e.}, $K=1$. 

The normalized generalization ability $\Gamma = \frac{1}{m} \text{rank} \ M_{\Gamma}$ is calculated using the same procedure, except that $M_{\Gamma}$ is formed by $m$ outputs from $m$ distinct inputs, each of which is preceded by the same string of constant bits. A system with perfect generalization has $\Gamma=1/m$, meaning that perturbations to the constant string of input bits do not cause a distinguishable change to the output.

An effective reservoir simultaneously maximizes $K$ and minimizes $\Gamma$, \textit{i.e.}, maximizes their difference $K - \Gamma \equiv \Delta$. Thus, $-1 < \Delta < 1$ is interpreted as the fraction of $L$ that is useful for computation. A perfect reservoir has $\Delta=1-1/m$, indicating that almost the entire consistency window provides useful information about the input, and $\Delta \leq 0$ represents a reservoir that fails to perform any useful information processing. We can therefore define an effective computational dimensionality $D=L\Delta$. Thus, the best reservoir is the one that maximizes $D$, which is bounded above by the length of the consistency window.

\begin{figure}[htb]
	\centering
		\includegraphics[scale=1]{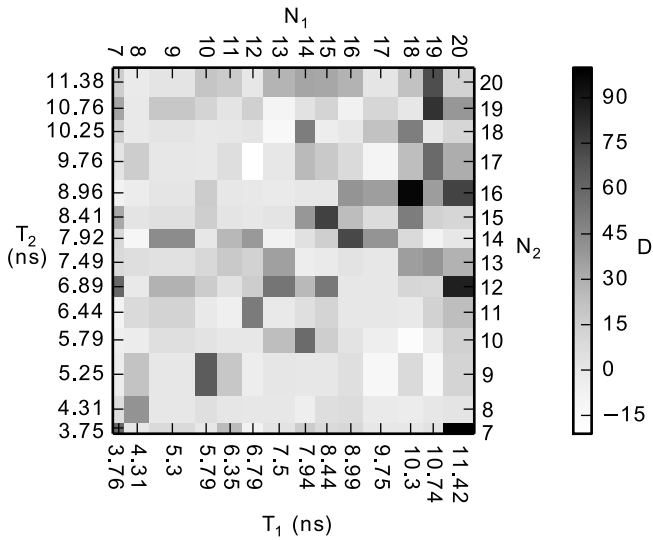}
	\caption{Dependence of effective computational dimensionality $D$ on the length of the delay lines. The tiles are different sizes due to the heterogeneity in the size of each delay element.}
	\label{fig4}
\end{figure}

We examine the effect of $T_1$ and $T_2$ on reservoir performance by measuring $D$ for different $N_1$ and $N_2$, shown in \cref{fig4}. The lack of symmetry about the diagonal suggests that the heterogeneity in delay times between lines 1 and 2, seen in \cref{tab1}, plays a significant role in determining the dynamics, and that the performance of the reservoir is highly dependent on the ratio $T_1/T_2$. This result is consistent with \cite{Dee84,Ghi85}, where it was shown that $T_1/T_2$ determines an upper bound on the period length of possible dynamics, giving rise to windows with extremely long periods. The parameters that maximize D are found to be $N_1 = 20$, $N_2 = 7$.

Using the parameters $N_1 = 20, N_2=7$ that maximize reservoir performance, we now demonstrate the reservoir's ability to identify input patterns. To train the reservoir, we collect the dynamics for the $2^n$ possible $n$-bit input patterns by saving $S$ samples of the output to on-chip RAM blocks at 400 MHz. Once the samples $x(t_0),x(t_1),...x(t_{S-1})$ have been collected, they are transferred to a PC and used to form linear combinations
\begin{equation}
C_i(t_j) = \sum_{k=0}^{W-1}{w_{k}^{ij} x(t_{j+k})}
\end{equation}
for each word $i=0,1,...,2^n-1$ beginning at each sample time $t_j=t_0,t_1,...t_{S-W-1}$. 

The linear combinations $C_i(t_j)$ are called classifiers, and the weights $w_{k}^{ij}$ that define them are computed by linear regression. The length of the classification window $W$ defines the number of samples used to compute each classifier. The target of classifier $C_i$ is 1 for the output pattern corresponding to input $i$ and $-1$ for all other patterns. Thus, when additional time series are provided to the $2^n$ classifiers, each computes a scalar ``score'' for every sample point. We regard successful classification as an event when the classifier with the largest score is the classifier corresponding to the actual input.

\begin{figure}[htb]
	\centering
		\includegraphics[scale=1]{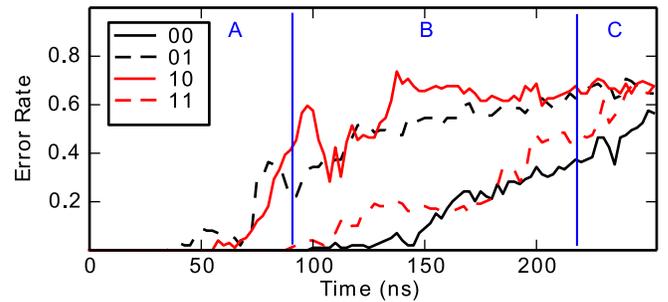}
	\caption{(color online) Word recognition error rate for 2-bit input sequences with $N_1 = 20$, $N_2 = 7$ elements. Solid blue (dark gray) vertical lines separate training regions A, B, and C, defined in the text.}
	\label{fig5}
\end{figure}

We present the classification performance of the reservoir in \cref{fig5}, which also demonstrates the effect of fading memory in the system. Error rates for each input word with $n=2$ bits, obtained by testing the classifiers with 100 time series for each input, are shown in \cref{fig5} for $N_1=20$, $N_2=7$ elements (the optimal parameters in \cref{fig4}). A time series length of $S=200$ samples (500 ns) is collected, and a classification window of $W=50$ samples (125 ns) is used. The consistency window for this system is measured to be 215 ns. Thus, there are three training regions. For classifiers $C_i(t_j)$ that begin at times $t_j$ less than 90 ns (region A), the entire window of classification lies within the region of consistency. Classifiers that begin at times greater than 215 ns (region C) are trained with only samples from outside of the consistency window. Classifiers starting between 90 and 215 ns (region B) are trained with a mix of data from inside and outside the consistency window.

The classification performance is the strongest in region A, with the error rate remaining below 10\% for approximately 70 ns, despite the input word being only 7.5 ns. In other words, the classifiers are able to faithfully reconstruct the reservoir's input state with reasonable accuracy even after the reservoir was no longer receiving information. Performance in region C plateaus at an error rate of $0.75$, \textit{i.e.}, chance, meaning that no useful information about the input remains in the output after the consistency window. Additionally, a rise in classification error can be observed throughout region B, indicating that the inclusion of data from outside the consistency window decreases performance. We observe a similar trend for words of $n = 3$ and $n = 4$ bits -- the error rate is lowest in region A, but gradually rises throughout region B to a plateau of $1-2^{-n}$ in region C. The error rate at a given time is generally higher for larger words, and the 32 possible words for $n=5$ bits cannot be accurately classified at any time. This behavior is similar to that found in a reservoir that used electrical recordings from neurons of the primary visual cortex of an anesthetized cat \cite{Nik09}.

We conclude that, despite its simplicity, the single time-delay feedback XOR reservoir produces dynamics suitable for RC. In addition, the reservoir is shown to perform a pattern recognition task. The complexity of the dynamics produced by the relatively simple system described in this article suggest that FPGAs are an exceptionally versatile platform for realizing physical RCs. 

Our results point to several future directions. For example, it is possible to build larger Boolean networks on FPGAs \cite{Ros13b,Ros14a,Ros14b}, which will allow for improved computation in two ways. First, recent work studying the computational ability of larger random Boolean networks without time delays \cite{Sny12,Sny13} motivates speculation that the performance of the number classification task described above can be improved with additional nodes. Second, the ability to run many parallel input nodes will allow a mix of spatial and temporal computation that could be attractive for speech and image recognition. The ability to train the reservoir using digital logic on the FGPA, instead of transmitting data to an external PC for post-processing, could also speed up training. As a result, the simple system studied here will be crucial to understanding the ability of much larger systems to process information.

N.D.H and D.J.G. gratefully acknowledge financial support by the U.S. Army Research Office through Grant $\#$W911NF-12-1-0099. N.D.H. also acknowledges the U.S. National Science Foundation for support through grant \#DGE-1068871. M.C.S and I.F. acknowledge support by MINECO (Spain) under Project TEC2012-36335 (TRIPHOP), and Govern de les Illes Balears via the program Grups Competitius.


\bibliographystyle{apsrev4-1}
\bibliography{Single_node_bib}

\begin{thebibliography}{24}%
\makeatletter
\providecommand \@ifxundefined [1]{%
 \@ifx{#1\undefined}
}%
\providecommand \@ifnum [1]{%
 \ifnum #1\expandafter \@firstoftwo
 \else \expandafter \@secondoftwo
 \fi
}%
\providecommand \@ifx [1]{%
 \ifx #1\expandafter \@firstoftwo
 \else \expandafter \@secondoftwo
 \fi
}%
\providecommand \natexlab [1]{#1}%
\providecommand \enquote  [1]{``#1''}%
\providecommand \bibnamefont  [1]{#1}%
\providecommand \bibfnamefont [1]{#1}%
\providecommand \citenamefont [1]{#1}%
\providecommand \href@noop [0]{\@secondoftwo}%
\providecommand \href [0]{\begingroup \@sanitize@url \@href}%
\providecommand \@href[1]{\@@startlink{#1}\@@href}%
\providecommand \@@href[1]{\endgroup#1\@@endlink}%
\providecommand \@sanitize@url [0]{\catcode `\\12\catcode `\$12\catcode
  `\&12\catcode `\#12\catcode `\^12\catcode `\_12\catcode `\%12\relax}%
\providecommand \@@startlink[1]{}%
\providecommand \@@endlink[0]{}%
\providecommand \url  [0]{\begingroup\@sanitize@url \@url }%
\providecommand \@url [1]{\endgroup\@href {#1}{\urlprefix }}%
\providecommand \urlprefix  [0]{URL }%
\providecommand \Eprint [0]{\href }%
\providecommand \doibase [0]{http://dx.doi.org/}%
\providecommand \selectlanguage [0]{\@gobble}%
\providecommand \bibinfo  [0]{\@secondoftwo}%
\providecommand \bibfield  [0]{\@secondoftwo}%
\providecommand \translation [1]{[#1]}%
\providecommand \BibitemOpen [0]{}%
\providecommand \bibitemStop [0]{}%
\providecommand \bibitemNoStop [0]{.\EOS\space}%
\providecommand \EOS [0]{\spacefactor3000\relax}%
\providecommand \BibitemShut  [1]{\csname bibitem#1\endcsname}%
\let\auto@bib@innerbib\@empty
\bibitem [{\citenamefont {Jaeger}(2001)}]{Jae01}%
  \BibitemOpen
  \bibfield  {author} {\bibinfo {author} {\bibfnamefont {H.}~\bibnamefont
  {Jaeger}},\ }\href@noop {} {\bibfield  {journal} {\bibinfo  {journal} {German
  National Research Center for Information Technology GMD Technical Report}\
  }\textbf {\bibinfo {volume} {148}},\ \bibinfo {pages} {34} (\bibinfo {year}
  {2001})}\BibitemShut {NoStop}%
\bibitem [{\citenamefont {Maass}\ \emph {et~al.}(2002)\citenamefont {Maass},
  \citenamefont {Natschl{\"a}ger},\ and\ \citenamefont {Markram}}]{Maa02}%
  \BibitemOpen
  \bibfield  {author} {\bibinfo {author} {\bibfnamefont {W.}~\bibnamefont
  {Maass}}, \bibinfo {author} {\bibfnamefont {T.}~\bibnamefont
  {Natschl{\"a}ger}}, \ and\ \bibinfo {author} {\bibfnamefont {H.}~\bibnamefont
  {Markram}},\ }\href@noop {} {\bibfield  {journal} {\bibinfo  {journal}
  {Neural Comput.}\ }\textbf {\bibinfo {volume} {14}},\ \bibinfo {pages} {2531}
  (\bibinfo {year} {2002})}\BibitemShut {NoStop}%
\bibitem [{\citenamefont {Luko{\v{s}}evi{\v{c}}ius}\ and\ \citenamefont
  {Jaeger}(2009)}]{Luk09}%
  \BibitemOpen
  \bibfield  {author} {\bibinfo {author} {\bibfnamefont {M.}~\bibnamefont
  {Luko{\v{s}}evi{\v{c}}ius}}\ and\ \bibinfo {author} {\bibfnamefont
  {H.}~\bibnamefont {Jaeger}},\ }\href@noop {} {\bibfield  {journal} {\bibinfo
  {journal} {Computer Science Review}\ }\textbf {\bibinfo {volume} {3}},\
  \bibinfo {pages} {127} (\bibinfo {year} {2009})}\BibitemShut {NoStop}%
\bibitem [{\citenamefont {Hammer}\ \emph {et~al.}(2009)\citenamefont {Hammer},
  \citenamefont {Schrauwen},\ and\ \citenamefont {Steil}}]{Ham09}%
  \BibitemOpen
  \bibfield  {author} {\bibinfo {author} {\bibfnamefont {B.}~\bibnamefont
  {Hammer}}, \bibinfo {author} {\bibfnamefont {B.}~\bibnamefont {Schrauwen}}, \
  and\ \bibinfo {author} {\bibfnamefont {J.~J.}\ \bibnamefont {Steil}},\ }in\
  \href@noop {} {\emph {\bibinfo {booktitle} {ESANN}}}\ (\bibinfo {year}
  {2009})\BibitemShut {NoStop}%
\bibitem [{\citenamefont {Appeltant}\ \emph {et~al.}(2011)\citenamefont
  {Appeltant}, \citenamefont {Soriano}, \citenamefont {Van~der Sande},
  \citenamefont {Danckaert}, \citenamefont {Massar}, \citenamefont {Dambre},
  \citenamefont {Schrauwen}, \citenamefont {Mirasso},\ and\ \citenamefont
  {Fischer}}]{App11}%
  \BibitemOpen
  \bibfield  {author} {\bibinfo {author} {\bibfnamefont {L.}~\bibnamefont
  {Appeltant}}, \bibinfo {author} {\bibfnamefont {M.~C.}\ \bibnamefont
  {Soriano}}, \bibinfo {author} {\bibfnamefont {G.}~\bibnamefont {Van~der
  Sande}}, \bibinfo {author} {\bibfnamefont {J.}~\bibnamefont {Danckaert}},
  \bibinfo {author} {\bibfnamefont {S.}~\bibnamefont {Massar}}, \bibinfo
  {author} {\bibfnamefont {J.}~\bibnamefont {Dambre}}, \bibinfo {author}
  {\bibfnamefont {B.}~\bibnamefont {Schrauwen}}, \bibinfo {author}
  {\bibfnamefont {C.~R.}\ \bibnamefont {Mirasso}}, \ and\ \bibinfo {author}
  {\bibfnamefont {I.}~\bibnamefont {Fischer}},\ }\href@noop {} {\bibfield
  {journal} {\bibinfo  {journal} {Nat. Commun.}\ }\textbf {\bibinfo {volume}
  {2}},\ \bibinfo {pages} {468} (\bibinfo {year} {2011})}\BibitemShut {NoStop}%
\bibitem [{\citenamefont {Larger}\ \emph {et~al.}(2012)\citenamefont {Larger},
  \citenamefont {Soriano}, \citenamefont {Brunner}, \citenamefont {Appeltant},
  \citenamefont {Guti{\'e}rrez}, \citenamefont {Pesquera}, \citenamefont
  {Mirasso},\ and\ \citenamefont {Fischer}}]{Lar12}%
  \BibitemOpen
  \bibfield  {author} {\bibinfo {author} {\bibfnamefont {L.}~\bibnamefont
  {Larger}}, \bibinfo {author} {\bibfnamefont {M.~C.}\ \bibnamefont {Soriano}},
  \bibinfo {author} {\bibfnamefont {D.}~\bibnamefont {Brunner}}, \bibinfo
  {author} {\bibfnamefont {L.}~\bibnamefont {Appeltant}}, \bibinfo {author}
  {\bibfnamefont {J.~M.}\ \bibnamefont {Guti{\'e}rrez}}, \bibinfo {author}
  {\bibfnamefont {L.}~\bibnamefont {Pesquera}}, \bibinfo {author}
  {\bibfnamefont {C.~R.}\ \bibnamefont {Mirasso}}, \ and\ \bibinfo {author}
  {\bibfnamefont {I.}~\bibnamefont {Fischer}},\ }\href@noop {} {\bibfield
  {journal} {\bibinfo  {journal} {Opt. Express}\ }\textbf {\bibinfo {volume}
  {20}},\ \bibinfo {pages} {3241} (\bibinfo {year} {2012})}\BibitemShut
  {NoStop}%
\bibitem [{\citenamefont {Paquot}\ \emph {et~al.}(2012)\citenamefont {Paquot},
  \citenamefont {Duport}, \citenamefont {Smerieri}, \citenamefont {Dambre},
  \citenamefont {Schrauwen}, \citenamefont {Haelterman},\ and\ \citenamefont
  {Massar}}]{Paq12}%
  \BibitemOpen
  \bibfield  {author} {\bibinfo {author} {\bibfnamefont {Y.}~\bibnamefont
  {Paquot}}, \bibinfo {author} {\bibfnamefont {F.}~\bibnamefont {Duport}},
  \bibinfo {author} {\bibfnamefont {A.}~\bibnamefont {Smerieri}}, \bibinfo
  {author} {\bibfnamefont {J.}~\bibnamefont {Dambre}}, \bibinfo {author}
  {\bibfnamefont {B.}~\bibnamefont {Schrauwen}}, \bibinfo {author}
  {\bibfnamefont {M.}~\bibnamefont {Haelterman}}, \ and\ \bibinfo {author}
  {\bibfnamefont {S.}~\bibnamefont {Massar}},\ }\href@noop {} {\bibfield
  {journal} {\bibinfo  {journal} {Sci. Rep.}\ }\textbf {\bibinfo {volume} {2}}
  (\bibinfo {year} {2012})}\BibitemShut {NoStop}%
\bibitem [{\citenamefont {Soriano}\ \emph {et~al.}(2013)\citenamefont
  {Soriano}, \citenamefont {Ort{\'\i}n}, \citenamefont {Brunner}, \citenamefont
  {Larger}, \citenamefont {Mirasso}, \citenamefont {Fischer},\ and\
  \citenamefont {Pesquera}}]{Sor13}%
  \BibitemOpen
  \bibfield  {author} {\bibinfo {author} {\bibfnamefont {M.~C.}\ \bibnamefont
  {Soriano}}, \bibinfo {author} {\bibfnamefont {S.}~\bibnamefont {Ort{\'\i}n}},
  \bibinfo {author} {\bibfnamefont {D.}~\bibnamefont {Brunner}}, \bibinfo
  {author} {\bibfnamefont {L.}~\bibnamefont {Larger}}, \bibinfo {author}
  {\bibfnamefont {C.~R.}\ \bibnamefont {Mirasso}}, \bibinfo {author}
  {\bibfnamefont {I.}~\bibnamefont {Fischer}}, \ and\ \bibinfo {author}
  {\bibfnamefont {L.}~\bibnamefont {Pesquera}},\ }\href@noop {} {\bibfield
  {journal} {\bibinfo  {journal} {Opt. Express}\ }\textbf {\bibinfo {volume}
  {21}},\ \bibinfo {pages} {12} (\bibinfo {year} {2013})}\BibitemShut {NoStop}%
\bibitem [{\citenamefont {Brunner}\ \emph {et~al.}(2013)\citenamefont
  {Brunner}, \citenamefont {Soriano}, \citenamefont {Mirasso},\ and\
  \citenamefont {Fischer}}]{Bru13}%
  \BibitemOpen
  \bibfield  {author} {\bibinfo {author} {\bibfnamefont {D.}~\bibnamefont
  {Brunner}}, \bibinfo {author} {\bibfnamefont {M.~C.}\ \bibnamefont
  {Soriano}}, \bibinfo {author} {\bibfnamefont {C.~R.}\ \bibnamefont
  {Mirasso}}, \ and\ \bibinfo {author} {\bibfnamefont {I.}~\bibnamefont
  {Fischer}},\ }\href@noop {} {\bibfield  {journal} {\bibinfo  {journal} {Nat.
  Commun.}\ }\textbf {\bibinfo {volume} {4}},\ \bibinfo {pages} {1364}
  (\bibinfo {year} {2013})}\BibitemShut {NoStop}%
\bibitem [{\citenamefont {Dejonckheere}\ \emph {et~al.}(2014)\citenamefont
  {Dejonckheere}, \citenamefont {Duport}, \citenamefont {Smerieri},
  \citenamefont {Fang}, \citenamefont {Oudar}, \citenamefont {Haelterman},\
  and\ \citenamefont {Massar}}]{Dej14}%
  \BibitemOpen
  \bibfield  {author} {\bibinfo {author} {\bibfnamefont {A.}~\bibnamefont
  {Dejonckheere}}, \bibinfo {author} {\bibfnamefont {F.}~\bibnamefont
  {Duport}}, \bibinfo {author} {\bibfnamefont {A.}~\bibnamefont {Smerieri}},
  \bibinfo {author} {\bibfnamefont {L.}~\bibnamefont {Fang}}, \bibinfo {author}
  {\bibfnamefont {J.}~\bibnamefont {Oudar}}, \bibinfo {author} {\bibfnamefont
  {M.}~\bibnamefont {Haelterman}}, \ and\ \bibinfo {author} {\bibfnamefont
  {S.}~\bibnamefont {Massar}},\ }\href@noop {} {\bibfield  {journal} {\bibinfo
  {journal} {Opt. Express}\ }\textbf {\bibinfo {volume} {22}},\ \bibinfo
  {pages} {10868} (\bibinfo {year} {2014})}\BibitemShut {NoStop}%
\bibitem [{\citenamefont {Ghil}\ and\ \citenamefont {Mullhaupt}(1985)}]{Ghi85}%
  \BibitemOpen
  \bibfield  {author} {\bibinfo {author} {\bibfnamefont {M.}~\bibnamefont
  {Ghil}}\ and\ \bibinfo {author} {\bibfnamefont {A.}~\bibnamefont
  {Mullhaupt}},\ }\href@noop {} {\bibfield  {journal} {\bibinfo  {journal} {J.
  Stat. Phys.}\ }\textbf {\bibinfo {volume} {41}},\ \bibinfo {pages} {125}
  (\bibinfo {year} {1985})}\BibitemShut {NoStop}%
\bibitem [{\citenamefont {Ghil}\ \emph {et~al.}(2008)\citenamefont {Ghil},
  \citenamefont {Zaliapin},\ and\ \citenamefont {Coluzzi}}]{Ghi08}%
  \BibitemOpen
  \bibfield  {author} {\bibinfo {author} {\bibfnamefont {M.}~\bibnamefont
  {Ghil}}, \bibinfo {author} {\bibfnamefont {I.}~\bibnamefont {Zaliapin}}, \
  and\ \bibinfo {author} {\bibfnamefont {B.}~\bibnamefont {Coluzzi}},\
  }\href@noop {} {\bibfield  {journal} {\bibinfo  {journal} {Physica D}\
  }\textbf {\bibinfo {volume} {237}},\ \bibinfo {pages} {2967} (\bibinfo {year}
  {2008})}\BibitemShut {NoStop}%
\bibitem [{\citenamefont {Uchida}\ \emph {et~al.}(2004)\citenamefont {Uchida},
  \citenamefont {McAllister},\ and\ \citenamefont {Roy}}]{Uch04}%
  \BibitemOpen
  \bibfield  {author} {\bibinfo {author} {\bibfnamefont {A.}~\bibnamefont
  {Uchida}}, \bibinfo {author} {\bibfnamefont {R.}~\bibnamefont {McAllister}},
  \ and\ \bibinfo {author} {\bibfnamefont {R.}~\bibnamefont {Roy}},\
  }\href@noop {} {\bibfield  {journal} {\bibinfo  {journal} {Phys. Rev. Lett.}\
  }\textbf {\bibinfo {volume} {93}},\ \bibinfo {pages} {244102} (\bibinfo
  {year} {2004})}\BibitemShut {NoStop}%
\bibitem [{\citenamefont {Rosin}\ \emph {et~al.}(2013)\citenamefont {Rosin},
  \citenamefont {Rontani}, \citenamefont {Gauthier},\ and\ \citenamefont
  {Sch{\"o}ll}}]{Ros13b}%
  \BibitemOpen
  \bibfield  {author} {\bibinfo {author} {\bibfnamefont {D.~P.}\ \bibnamefont
  {Rosin}}, \bibinfo {author} {\bibfnamefont {D.}~\bibnamefont {Rontani}},
  \bibinfo {author} {\bibfnamefont {D.~J.}\ \bibnamefont {Gauthier}}, \ and\
  \bibinfo {author} {\bibfnamefont {E.}~\bibnamefont {Sch{\"o}ll}},\
  }\href@noop {} {\bibfield  {journal} {\bibinfo  {journal} {Chaos}\ }\textbf
  {\bibinfo {volume} {23}},\ \bibinfo {pages} {025102} (\bibinfo {year}
  {2013})}\BibitemShut {NoStop}%
\bibitem [{\citenamefont {Cavalcante}\ \emph {et~al.}(2010)\citenamefont
  {Cavalcante}, \citenamefont {Gauthier}, \citenamefont {Socolar},\ and\
  \citenamefont {Zhang}}]{Cav10}%
  \BibitemOpen
  \bibfield  {author} {\bibinfo {author} {\bibfnamefont {H.~L. D. d.~S.}\
  \bibnamefont {Cavalcante}}, \bibinfo {author} {\bibfnamefont {D.~J.}\
  \bibnamefont {Gauthier}}, \bibinfo {author} {\bibfnamefont {J.~E.~S.}\
  \bibnamefont {Socolar}}, \ and\ \bibinfo {author} {\bibfnamefont
  {R.}~\bibnamefont {Zhang}},\ }\href@noop {} {\bibfield  {journal} {\bibinfo
  {journal} {Philos. Tr. R. Soc. A}\ }\textbf {\bibinfo {volume} {368}},\
  \bibinfo {pages} {495} (\bibinfo {year} {2010})}\BibitemShut {NoStop}%
\bibitem [{\citenamefont {Zhang}\ \emph {et~al.}(2009)\citenamefont {Zhang},
  \citenamefont {Cavalcante}, \citenamefont {Gao}, \citenamefont {Gauthier},
  \citenamefont {Socolar}, \citenamefont {Adams},\ and\ \citenamefont
  {Lathrop}}]{Zha09}%
  \BibitemOpen
  \bibfield  {author} {\bibinfo {author} {\bibfnamefont {R.}~\bibnamefont
  {Zhang}}, \bibinfo {author} {\bibfnamefont {H.~L. D. d.~S.}\ \bibnamefont
  {Cavalcante}}, \bibinfo {author} {\bibfnamefont {Z.}~\bibnamefont {Gao}},
  \bibinfo {author} {\bibfnamefont {D.~J.}\ \bibnamefont {Gauthier}}, \bibinfo
  {author} {\bibfnamefont {J.~E.~S.}\ \bibnamefont {Socolar}}, \bibinfo
  {author} {\bibfnamefont {M.~M.}\ \bibnamefont {Adams}}, \ and\ \bibinfo
  {author} {\bibfnamefont {D.~P.}\ \bibnamefont {Lathrop}},\ }\href@noop {}
  {\bibfield  {journal} {\bibinfo  {journal} {Phys. Rev. E}\ }\textbf {\bibinfo
  {volume} {80}},\ \bibinfo {pages} {045202} (\bibinfo {year}
  {2009})}\BibitemShut {NoStop}%
\bibitem [{\citenamefont {Legenstein}\ and\ \citenamefont
  {Maass}(2007)}]{Leg07}%
  \BibitemOpen
  \bibfield  {author} {\bibinfo {author} {\bibfnamefont {R.}~\bibnamefont
  {Legenstein}}\ and\ \bibinfo {author} {\bibfnamefont {W.}~\bibnamefont
  {Maass}},\ }\href@noop {} {\bibfield  {journal} {\bibinfo  {journal} {Neural
  Networks}\ }\textbf {\bibinfo {volume} {20}},\ \bibinfo {pages} {323}
  (\bibinfo {year} {2007})}\BibitemShut {NoStop}%
\bibitem [{\citenamefont {B{\"u}sing}\ \emph {et~al.}(2010)\citenamefont
  {B{\"u}sing}, \citenamefont {Schrauwen},\ and\ \citenamefont
  {Legenstein}}]{Bus10}%
  \BibitemOpen
  \bibfield  {author} {\bibinfo {author} {\bibfnamefont {L.}~\bibnamefont
  {B{\"u}sing}}, \bibinfo {author} {\bibfnamefont {B.}~\bibnamefont
  {Schrauwen}}, \ and\ \bibinfo {author} {\bibfnamefont {R.}~\bibnamefont
  {Legenstein}},\ }\href@noop {} {\bibfield  {journal} {\bibinfo  {journal}
  {Neural Comput.}\ }\textbf {\bibinfo {volume} {22}},\ \bibinfo {pages} {1272}
  (\bibinfo {year} {2010})}\BibitemShut {NoStop}%
\bibitem [{\citenamefont {Dee}\ and\ \citenamefont {Ghil}(1984)}]{Dee84}%
  \BibitemOpen
  \bibfield  {author} {\bibinfo {author} {\bibfnamefont {D.}~\bibnamefont
  {Dee}}\ and\ \bibinfo {author} {\bibfnamefont {M.}~\bibnamefont {Ghil}},\
  }\href@noop {} {\bibfield  {journal} {\bibinfo  {journal} {SIAM J. Appl.
  Math.}\ }\textbf {\bibinfo {volume} {44}},\ \bibinfo {pages} {111} (\bibinfo
  {year} {1984})}\BibitemShut {NoStop}%
\bibitem [{\citenamefont {Nikoli{\'c}}\ \emph {et~al.}(2009)\citenamefont
  {Nikoli{\'c}}, \citenamefont {H{\"a}usler}, \citenamefont {Singer},\ and\
  \citenamefont {Maass}}]{Nik09}%
  \BibitemOpen
  \bibfield  {author} {\bibinfo {author} {\bibfnamefont {D.}~\bibnamefont
  {Nikoli{\'c}}}, \bibinfo {author} {\bibfnamefont {S.}~\bibnamefont
  {H{\"a}usler}}, \bibinfo {author} {\bibfnamefont {W.}~\bibnamefont {Singer}},
  \ and\ \bibinfo {author} {\bibfnamefont {W.}~\bibnamefont {Maass}},\
  }\href@noop {} {\bibfield  {journal} {\bibinfo  {journal} {PLoS Biol.}\
  }\textbf {\bibinfo {volume} {7}},\ \bibinfo {pages} {e1000260} (\bibinfo
  {year} {2009})}\BibitemShut {NoStop}%
\bibitem [{\citenamefont {Rosin}\ \emph
  {et~al.}(2014{\natexlab{a}})\citenamefont {Rosin}, \citenamefont {Rontani},\
  and\ \citenamefont {Gauthier}}]{Ros14a}%
  \BibitemOpen
  \bibfield  {author} {\bibinfo {author} {\bibfnamefont {D.~P.}\ \bibnamefont
  {Rosin}}, \bibinfo {author} {\bibfnamefont {D.}~\bibnamefont {Rontani}}, \
  and\ \bibinfo {author} {\bibfnamefont {D.~J.}\ \bibnamefont {Gauthier}},\
  }\href@noop {} {\bibfield  {journal} {\bibinfo  {journal} {Phys. Rev. E}\
  }\textbf {\bibinfo {volume} {89}},\ \bibinfo {pages} {042907} (\bibinfo
  {year} {2014}{\natexlab{a}})}\BibitemShut {NoStop}%
\bibitem [{\citenamefont {Rosin}\ \emph
  {et~al.}(2014{\natexlab{b}})\citenamefont {Rosin}, \citenamefont {Rontani},
  \citenamefont {Haynes}, \citenamefont {Sch\"oll},\ and\ \citenamefont
  {Gauthier}}]{Ros14b}%
  \BibitemOpen
  \bibfield  {author} {\bibinfo {author} {\bibfnamefont {D.~P.}\ \bibnamefont
  {Rosin}}, \bibinfo {author} {\bibfnamefont {D.}~\bibnamefont {Rontani}},
  \bibinfo {author} {\bibfnamefont {N.~D.}\ \bibnamefont {Haynes}}, \bibinfo
  {author} {\bibfnamefont {E.}~\bibnamefont {Sch\"oll}}, \ and\ \bibinfo
  {author} {\bibfnamefont {D.~J.}\ \bibnamefont {Gauthier}},\ }\href {\doibase
  10.1103/PhysRevE.90.030902} {\bibfield  {journal} {\bibinfo  {journal} {Phys.
  Rev. E}\ }\textbf {\bibinfo {volume} {90}},\ \bibinfo {pages} {030902}
  (\bibinfo {year} {2014}{\natexlab{b}})}\BibitemShut {NoStop}%
\bibitem [{\citenamefont {Snyder}\ \emph {et~al.}(2012)\citenamefont {Snyder},
  \citenamefont {Goudarzi},\ and\ \citenamefont {Teuscher}}]{Sny12}%
  \BibitemOpen
  \bibfield  {author} {\bibinfo {author} {\bibfnamefont {D.}~\bibnamefont
  {Snyder}}, \bibinfo {author} {\bibfnamefont {A.}~\bibnamefont {Goudarzi}}, \
  and\ \bibinfo {author} {\bibfnamefont {C.}~\bibnamefont {Teuscher}},\ }in\
  \href@noop {} {\emph {\bibinfo {booktitle} {Artificial Life}}},\
  Vol.~\bibinfo {volume} {13}\ (\bibinfo {year} {2012})\ pp.\ \bibinfo {pages}
  {259--266}\BibitemShut {NoStop}%
\bibitem [{\citenamefont {Snyder}\ \emph {et~al.}(2013)\citenamefont {Snyder},
  \citenamefont {Goudarzi},\ and\ \citenamefont {Teuscher}}]{Sny13}%
  \BibitemOpen
  \bibfield  {author} {\bibinfo {author} {\bibfnamefont {D.}~\bibnamefont
  {Snyder}}, \bibinfo {author} {\bibfnamefont {A.}~\bibnamefont {Goudarzi}}, \
  and\ \bibinfo {author} {\bibfnamefont {C.}~\bibnamefont {Teuscher}},\
  }\href@noop {} {\bibfield  {journal} {\bibinfo  {journal} {Phys. Rev. E}\
  }\textbf {\bibinfo {volume} {87}},\ \bibinfo {pages} {042808} (\bibinfo
  {year} {2013})}\BibitemShut {NoStop}%
\end{thebibliography}%

\end{document}